\documentclass[twocolumn]{revtex4-1}
\usepackage[latin9]{inputenc}
\usepackage{amsmath}
\usepackage{amstext}
\usepackage{graphicx}
\usepackage{physics}
\begin{document}
\title{Limits on the Stochastic Gravitational Wave Background and Prospects for Single Source Detection with GRACE Follow-On}
\author{M.P. Ross}
\email[]{mpross2@uw.edu}
\author{C.A. Hagedorn}
\author{E.A. Shaw}
\author{A.L. Lockwood}
\author{B.M. Iritani}
\author{J.G. Lee}
\author{K. Venkateswara}
\author{J. H. Gundlach}
\affiliation{Center for Experimental Nuclear Physics and Astrophysics, University of Washington, Seattle, Washington,
98195, USA}

\begin{abstract}
With a reinterpretation of recent results, the GRACE Follow-On mission can be applied to gravitational wave astronomy. Existing GRACE Follow-On data constrain the stochastic gravitational wave background to $\Omega_{GW}<3.3\times10^{7}$ at 100 mHz. With a dedicated analysis, GRACE Follow-On may be able to detect the inspiral of local neutron star binaries, inspiral of subgalactic stellar-mass black hole binaries, or mergers of intermediate-mass black hole binaries within the Milky Way. 
\end{abstract}

\maketitle

\section{Introduction}
GRACE Follow-On (GRACE-FO) \cite{GRACE} is a satellite mission currently in Earth orbit studying the terrestrial gravitational field. We show that by interpreting recent results from this mission as strain, one can (1) constrain the gravitational wave background and (2) provide an opportunity for single-source gravitational wave (GW) searches. 

Incoherent GW must be present throughout the
universe at some amplitude. 
A broadband stochastic background is one of the few observable consequences of
cosmic inflation \cite{inflation} and a signature of proposed exotic
phenomena such as cosmic strings \cite{strings}. Measurements of,
and constraints on, such a background would yield insight into of
a wide range of phenomena, from inflationary models and predictions
of string theory to populations of black holes and neutron stars. Existing limits on the stochastic background come from pulsar timing \cite{Pulsar}, satellite ranging \cite{Cassini, Uly}, torsion balances \cite{TOBA, TOBA2}, terrestrial interferometers \cite{LIGOStoch}, and geophysical observations \cite{earthModes, Seismic, Lunar}. 

The most prominent single-source GW are emitted either by binary systems or spinning asymmetric stars. While asymmetric stars emit GW with a quasi-stationary frequency, inspiraling binary systems radiate semi-monotone GW before a rapid merger and ring-down phase. The only direct observations of GW have been from the merger of black hole binaries and binary neutron stars \citep{GWTC, 190425, 190412} by the LIGO \citep{LIGOgw} and Virgo \citep{Virgo} observatories.

\section{Strain Measurements}

As a GW passes two inertial masses nominally
separated by a distance $x_{0}$ their relative displacement, $x$,
is \cite{maggiore2008gravitational}: 

\begin{centering}
\begin{equation}
x(t)=x_{0}h_{xx}\cos\left(2\pi ft+\psi\right)
\end{equation}
\end{centering}

\begin{align}
h_{xx}=&\ h_+\ (\cos^2\theta\ \cos^2\phi-\sin^2\phi ) \label{GWeq} \\
\nonumber&+ 2 h_\times\ \cos\theta\ \sin \phi\ \cos\phi
\end{align}
where $h_{+,\times}$ are the plus and cross-polarization components of the gravitational
wave, $\theta$ and $\phi$ are the polar and azimuthal angles of the source, $f$ is the wave frequency, and $\psi$
is an arbitrary phase. Here, the plus and cross polarizations are defined in the source frame.

\begin{figure}[!h]
\centering \includegraphics[width=0.5\textwidth]{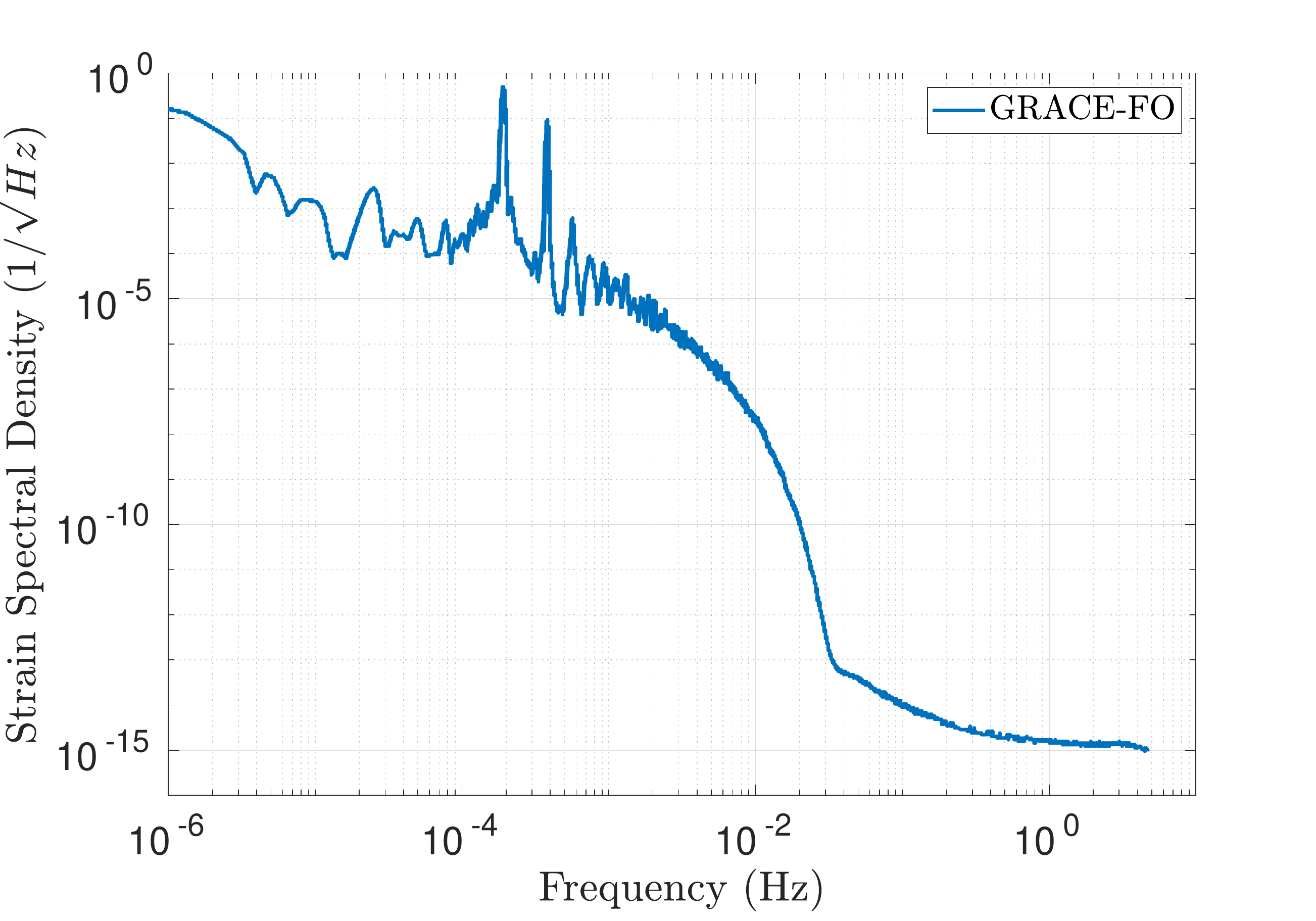}
\caption{Strain amplitude spectral density for the GRACE-FO mission. The increase in apparent noise below 30 mHz is due to terrestrial gravitational signals and not instrumental noise. }
\label{Strain} 
\end{figure}

The GRACE-FO satellite mission has produced
sensitive relative-displacement measurements of an inertial mass
pair formed by its two satellites separated
by $x_{0}=220$~km. The published displacement
spectrum \cite{GRACE} can be converted to strain via $\tilde{h}_{xx}(f)=\tilde{x}(f)/x_{0}$.
The inferred strain-noise spectral density is shown in Figure~\ref{Strain}. As noted in Ref.~\cite{GRACE}, the noise below 30 mHz may be significantly reduced by subtracting models of the terrestrial gravitational field.

\begin{widetext}

\begin{figure}[!h]
\centering \includegraphics[width=1\textwidth]{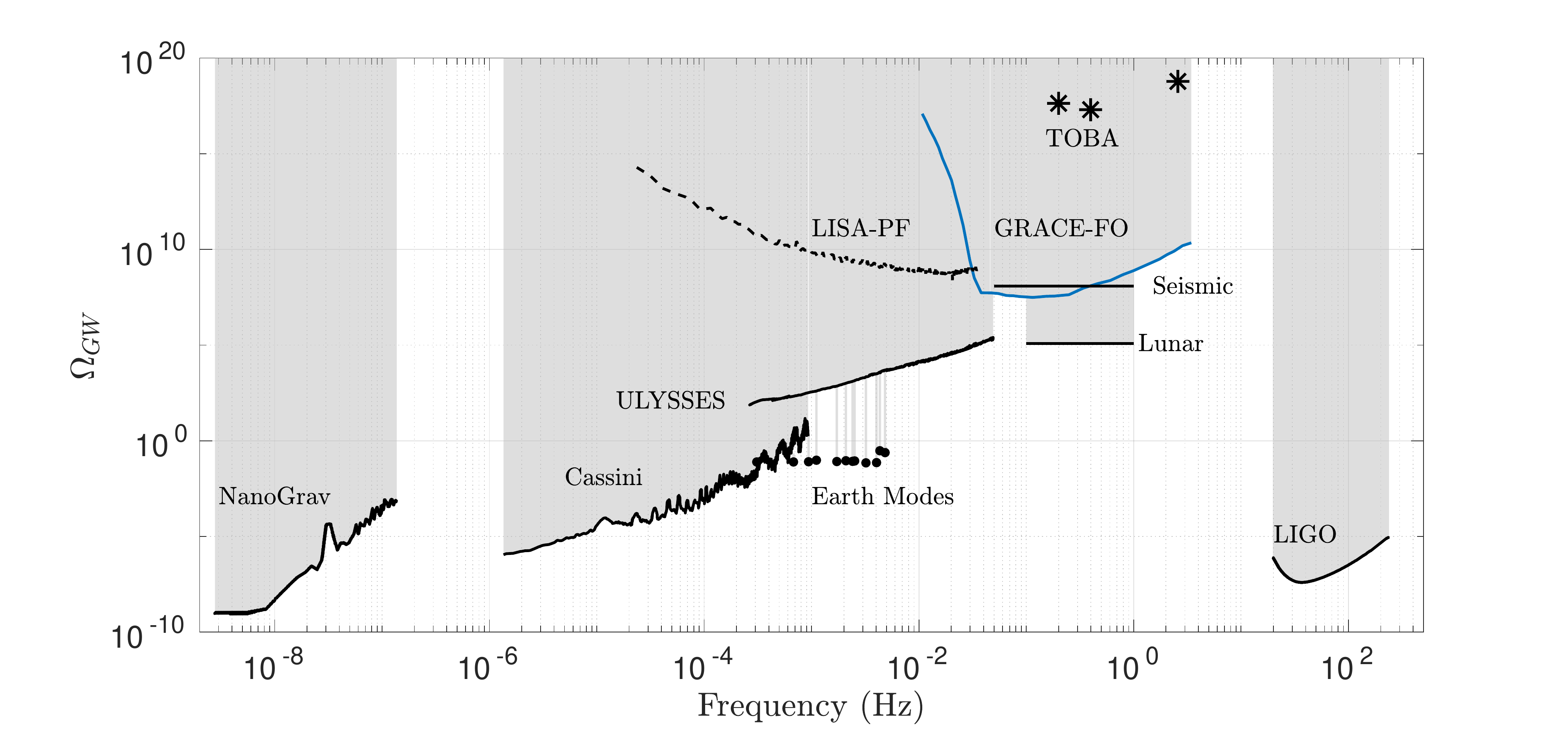} \caption{Limits on the stochastic gravitational wave background set by
GRACE-FO (blue) and LISA-PF (dashed). Analysis of LISA-PF is detailed in Appendix B. Additionally, limits set
by NanoGrav \cite{Pulsar}, Cassini \cite{Cassini}, ULYSSES \cite{Uly}, TOBA \cite{TOBA,TOBA2}, LIGO \cite{LIGOStoch}, observations
of the earth's normal modes \cite{earthModes}, terrestrial seismic motion \cite{Seismic}, and lunar seismic motion \cite{Lunar} are shown. Gray shading indicates regions excluded at 95\% confidence.}
\label{GW} 
\end{figure}

\end{widetext}

\section{Background Constraints}

Assuming an isotropic, unpolarized, stochastic gravitational wave
background, this strain noise can be interpreted as limits on the cosmological
energy density of gravitational waves, $\Omega_{\text{GW}}$ \cite{theory}, by integrating
Equation~\ref{GWeq} over incidence angle and polarization. 
\begin{equation}
\Omega_{\text{\text{GW}}}(f)=\frac{f}{\rho_{\text{crit}}}\dv{\rho_{\text{GW}}}{f}=\frac{32\pi^{3}}{3H_{0}^{2}}f^{3}\frac{15}{4}\tilde{h}_{\text{meas}}(f)^{2}\label{background}
\end{equation}

where $\rho_{\text{crit}}$ is the critical energy density of the universe, $\rho_{\text{GW}}$ is the energy density of gravitational waves, $H_{0}$ is Hubble constant, and $h_{\text{meas}}$ is the measured displacement interpreted as strain. A Hubble constant of $H_0=70.3\ \text{km}\ \text{s}^{-1}\ \text{Mpc}^{-1}$ was used \cite{hubble}. The factor of $15/4$ corrects for the polarization and angular sensitivity of the instruments in question, see Appendix A. 

Confidence intervals were extracted from the GRACE-FO strain spectrum by separating the data into frequency bins which each encompassed 20 data points. Limits for each bin were set at the 95th percentile to yield the results shown in Figure~\ref{GW}.
We anticipate that a dedicated analysis would improve these constraints further.

\section{Single-Sources}

In addition to setting stochastic GW limits,
the GRACE-FO mission may allow for gravitational wave searches at frequencies between 30 mHz and 5 Hz. Both stellar-mass black hole binaries and neutron star
binaries emit GW in the band of interest during their 
inspiral phase, while intermediate-mass black hole binaries would merge within the band.

Due to the orbit of the satellites, a GW seen by GRACE-FO would be modulated as the antenna pattern sweeps across the sky. This modulation can be approximated by rotating $\phi$ at the orbit frequency, $ f_{\text{orb}}=1/94.5\ \text{min}$ \cite{GRACE}. 
\begin{equation}
\phi(t)=2\pi  f_{\text{orb}} t+\phi_0
\end{equation}

Simulated antenna factors (the angular dependence of Equation~\ref{GWeq}) for GRACE-FO, at a collection of $\theta$ values, are shown in Figure \ref{Sim}. For sources in-line with the orbital axis ($\theta=0$) the instrument's polarization sensitivity is modulated, while for sources in the orbital plane ($\theta=\pi/2$) the total observed power is modulated.

Detailed analysis of the orbit
of the satellites and backgrounds due to the terrestrial gravitational field would be required for a definitive search. However, an estimate of the characteristic strain \cite{charStr} sensitivity can be
inferred from the strain spectral density. This
is shown in Figure \ref{Trans} along with a collection of known
and speculative sources. The best characteristic strain sensitivity is $2 \times 10^{-15}$ at 0.5 Hz.

\pagebreak

\begin{widetext}

\begin{figure}[!h]
\centering \includegraphics[width=1\textwidth]{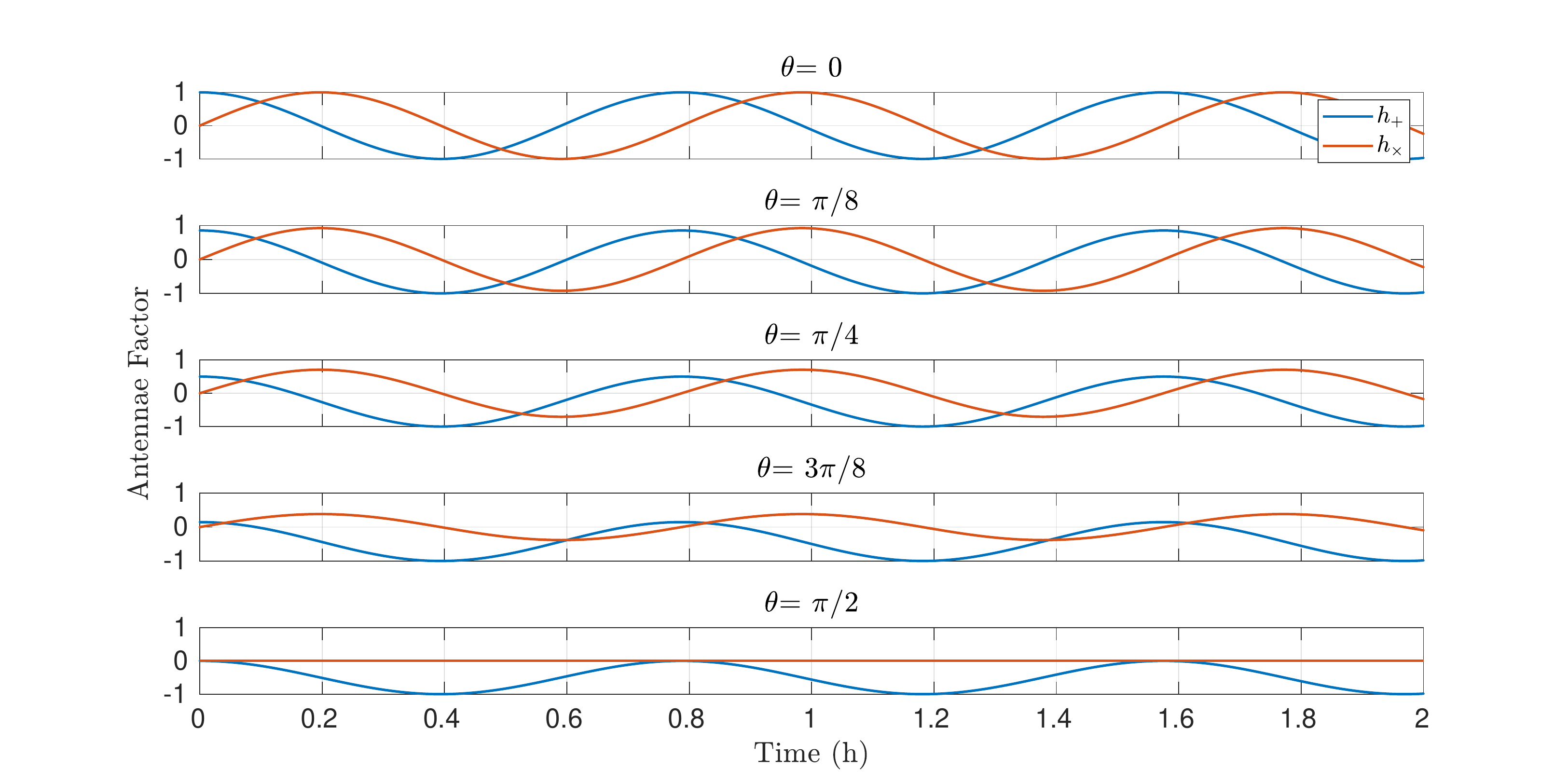} \caption{Simulated antenna factors for $h_+$ and $h_\times$ where $\theta$ is the polar angle of the source. For sources in-line with the satellite's orbital axis ($\theta=0$) the instrument's polarization sensitivity is modulated by the orbit, while for sources in the orbital plane ($\theta=\pi/2$) the total observed power is modulated.}
\label{Sim} 
\end{figure}

\begin{figure}[!h]
\centering \includegraphics[width=1\textwidth]{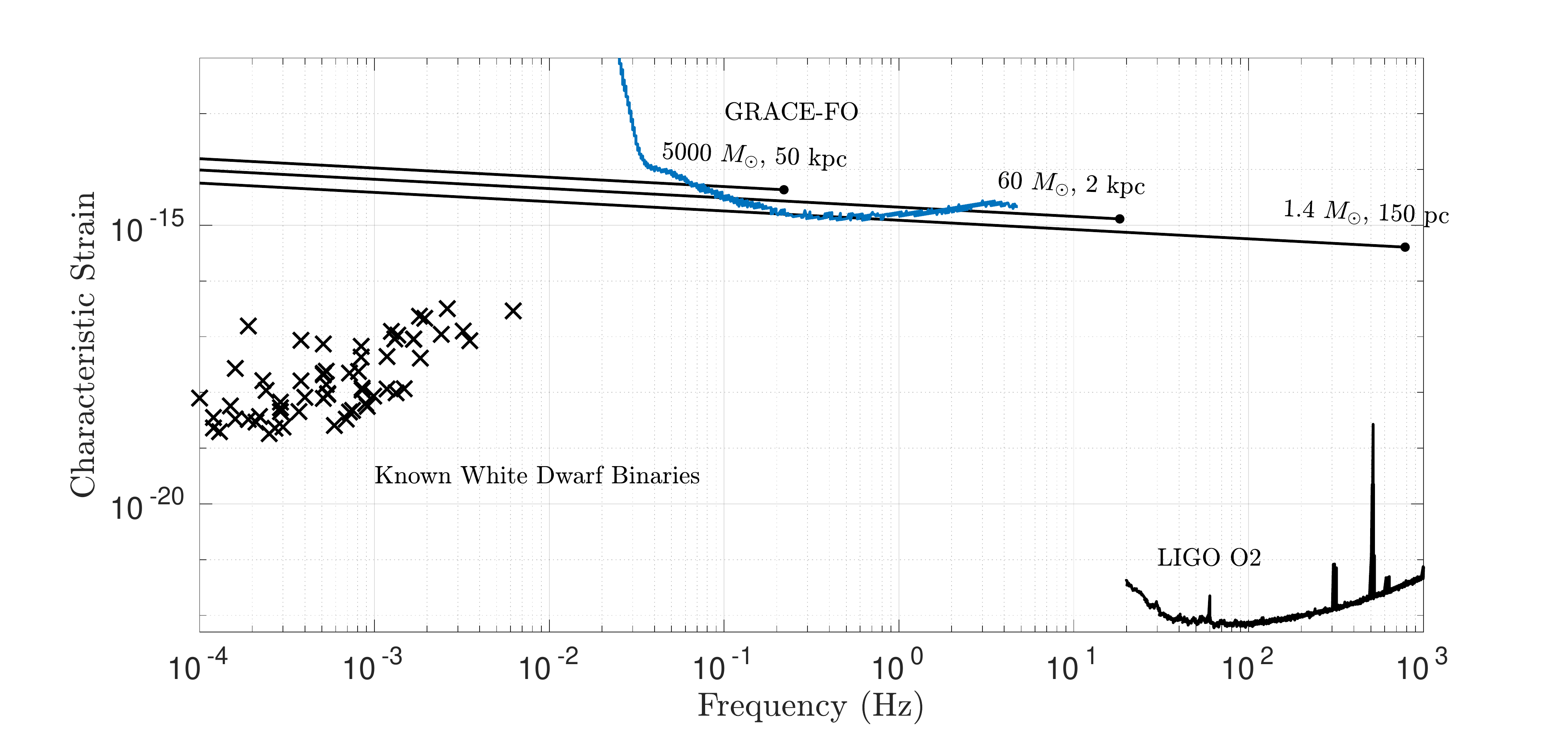} \caption{Estimates of the characteristic strain sensitivity of GRACE-FO  along with the expected strain from known white dwarf binaries \cite{whiteDwarf, Burdge_2019, Hermes_2012, burdge2019general} and a sensitivity curve for LIGO's second observing run \cite{LIGOStrain}. Also shown are lines that correspond to the predicted characteristic strain of an equal component-mass 1.4 $M_\odot$ binary at 150 pc, a 60 $M_\odot$ at 2 kpc, and a 5000 $M_\odot$ at 50 kpc \cite{charStr}. The points at the end of each line note the merger frequency for the corresponding system. These approximations are only valid for the inspiral phase of binary evolution.}\label{Trans} 
\end{figure}

\end{widetext}

\pagebreak

\begin{figure}[!h]
\centering \includegraphics[width=0.5\textwidth]{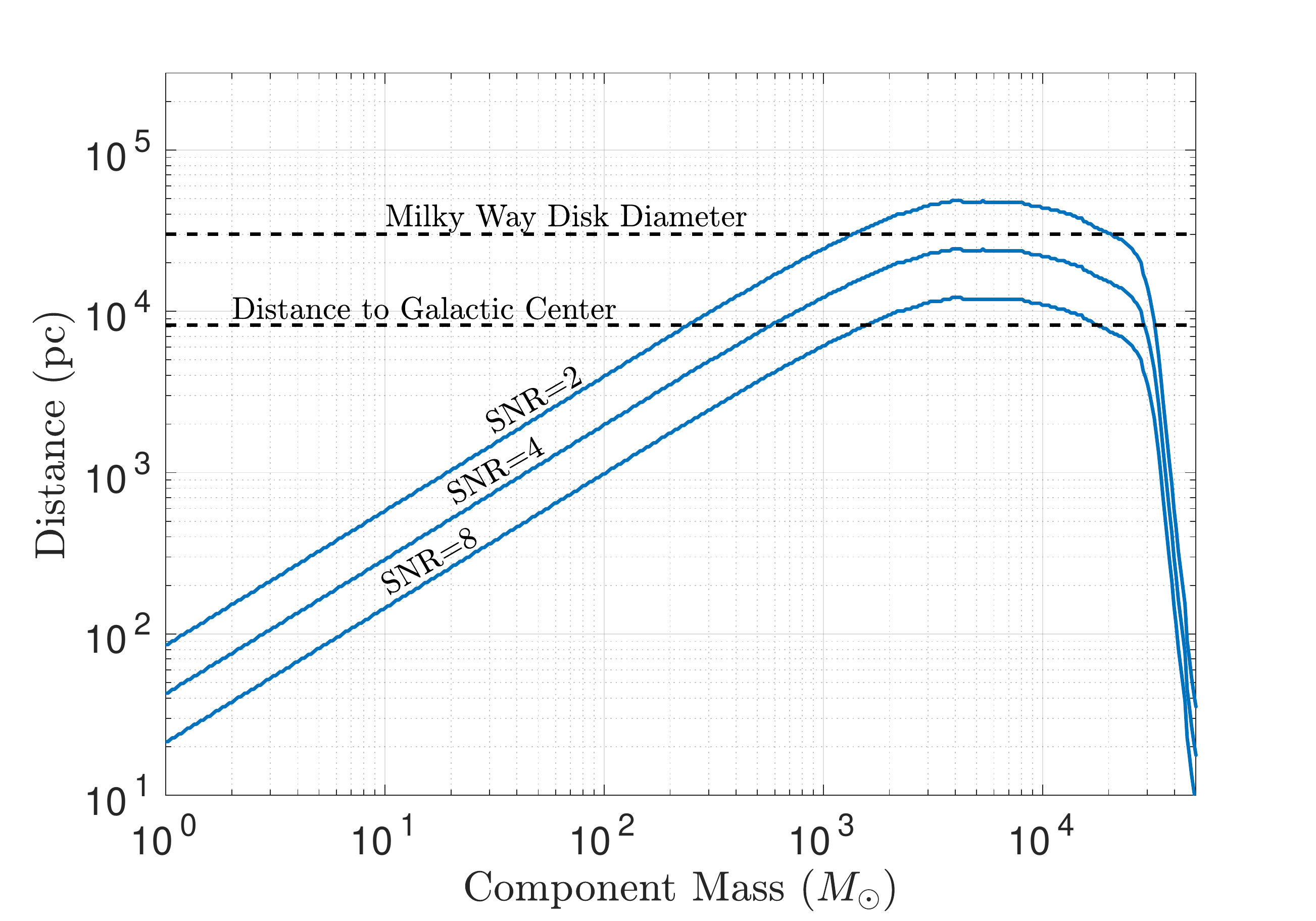} \caption{Maximum detectable distance for equal component-mass binaries with SNR thresholds of 2, 4, and 8. For systems with component-mass of 2000-10000 $M_\odot$ the detection volume encompasses most of the Milky Way.}
\label{ViewDist} 
\end{figure}
\hspace{\textwidth}

\hspace{\textwidth}



The characteristic strain allows for the estimation of the signal-to-noise ratio (SNR) of an optimal search for a given source \cite{charStr}. Figure \ref{ViewDist} shows the estimated maximum detectable distance for a given equal component-mass system with a select SNR threshold. 

The detection range peaks at component-masses of 2000-10000 $M_\odot$ with a volume encompassing most of the Milky Way. Above 10000 $M_\odot$, terrestrial gravitational signals severely limit detection capability while below $\sim100\ M_\odot$ the decreased emission does not provide an appreciable detection volume.

\section{Conclusion}

Existing GRACE Follow-On data constrain the stochastic gravitational
wave background to levels of $\Omega_{GW}<3.3\times10^{7}$ at 100 mHz. These constraints are many orders of magnitude
from expected sources \cite{inflation}, yet provide an independent set of limits in a frequency range where existing limits are sparse.

With a dedicated analysis, GRACE-FO is capable of single-source
GW searches. Promising target systems include the inspiral of local neutron star binaries or subgalactic stellar-mass black hole binaries, and the mergers of intermediate-mass black hole binaries within the Milky Way. 

Although the application of this mission to gravitational wave astronomy will be supplanted by the launch of LISA \cite{LISAFuture}, GRACE-FO has the capability to yield opportunistic insight into the local universe.

\hspace{\textwidth}

\begin{acknowledgements}
The authors would like to thank Eric Adelberger for his invaluable input and the anonymous referees for their insightful suggestions. This work was supported by funding from the NSF under Awards PHY-1607385, PHY-1607391, PHY-1912380 and PHY-1912514.
\end{acknowledgements}

 \bibliographystyle{unsrtnat}
\bibliography{SatStochastic.bib}

\section{Appendix A}\label{appA}
The strain along a single axis caused by a GW is \cite{maggiore2008gravitational}:
\begin{align}
h_{xx}&=\ h_+\ (\cos^2\theta\ \cos^2\phi-\sin^2\phi )\\
\nonumber &+ 2 h_\times\ \cos\theta\ \sin \phi\ \cos\phi 
\end{align}
where $h_{xx}$ is the strain along the x-axis, $h_{+,\times}$ are the strain amplitudes respectively in the plus and cross polarization, and $\theta$ and $\phi$ are the polar and azimuth angles, respectively.

Assuming isotropic emission, the fraction of incident power that is captured by a single-axis instrument can be found by:
\begin{align}
\tilde{h}_{\text{meas}}^2(f)&=\frac{1}{4\pi} \int \tilde{h}_{xx}^2(f) d\Omega \\
\nonumber &=\frac{1}{4\pi} \int d\Omega \big[\tilde{h}_+^2(f)\ (\cos^2\theta\ \cos^2\phi-\sin^2\phi )^2 \\
\ & \quad \quad + 4 \tilde{h}_\times^2(f)\ (\cos\theta\ \sin \phi\ \cos\phi)^2 \\
\nonumber & \quad \quad+2\ \tilde{h}_+(f)\ \tilde{h}_\times(f)\ (\cos^2\theta\ \cos^2\phi-\sin^2\phi )\ \\
\nonumber & \quad \quad  \quad \times\cos\theta\ \sin \phi\ \cos\phi\big]\\
 &=\frac{1}{4\pi} \bigg(\tilde{h}_+^2(f)\ \frac{22\pi}{15}+ 4 \tilde{h}_\times^2(f)\ \frac{\pi}{6} \bigg)
\end{align}

Further assuming an unpolarized background, the polarization can be averaged to yield:
\begin{align}
\tilde{h}_+^2(f)&=\tilde{h}_\times^2(f)=\frac{1}{2}\tilde{h}^2(f) \\
\tilde{h}_{\text{meas}}^2(f)&=\frac{1}{8\pi} \tilde{h}^2(f) \bigg(\frac{22\pi}{15}+  \frac{2\pi}{3} \bigg) \\
\tilde{h}_{\text{meas}}^2(f)&=\frac{4}{15} \tilde{h}^2(f)
\end{align}

\pagebreak

\section{Appendix B}\label{appB}

\begin{figure}[!h]
\centering \includegraphics[width=0.44\textwidth]{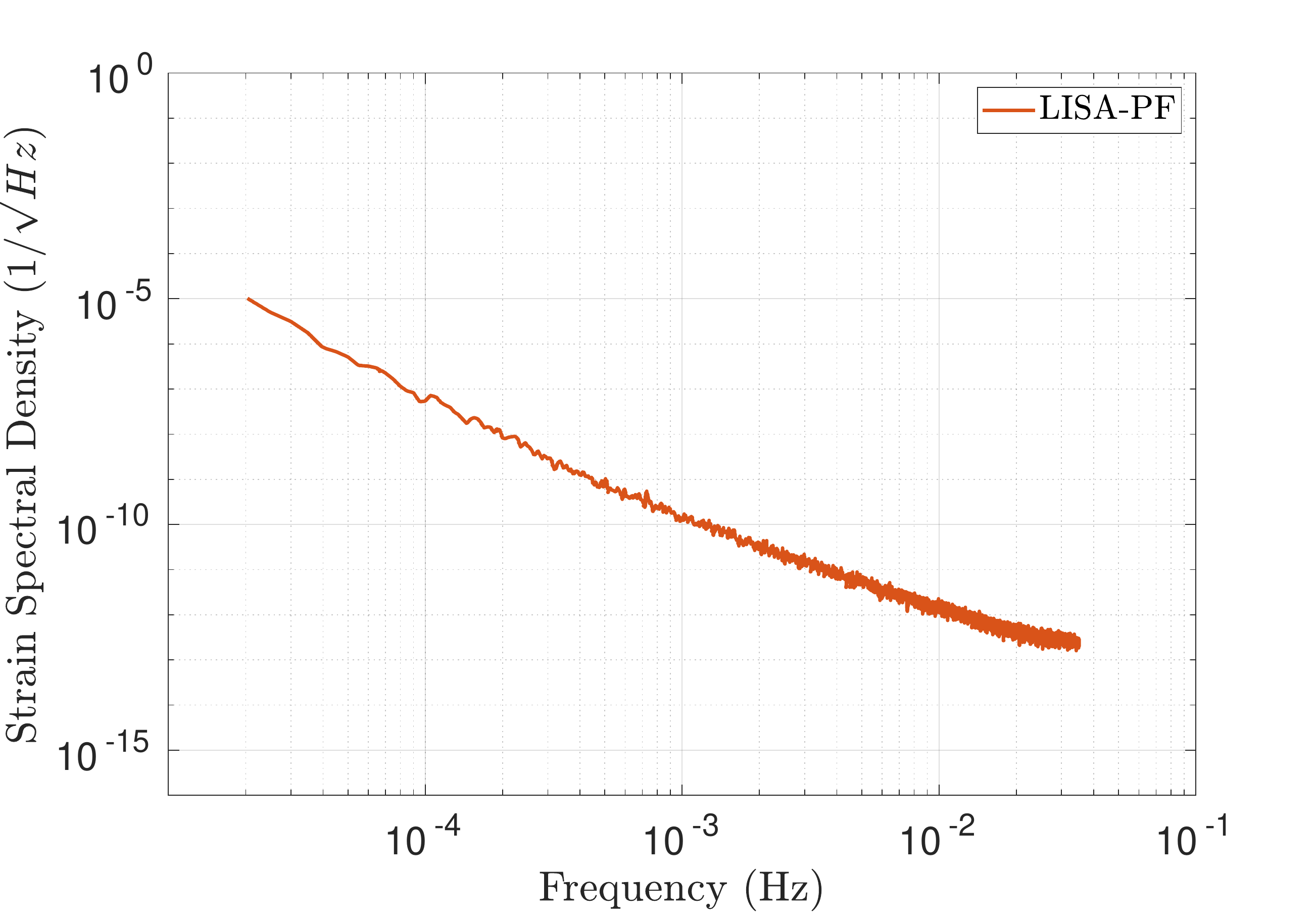}
\caption{Strain amplitude spectral density for LISA-PF.}
\label{StrainLPF} 
\end{figure}

The technique used with GRACE-FO can also be applied to LISA Pathfinder (LISA-PF) \cite{LISA} which was flown to demonstrate the drag-free technology needed for the future LISA mission \cite{LISAFuture}. For LISA-PF the mass-pair is formed by the proof masses, separated by $x_{0}=376$ ~mm.  LISA-PF has published an acceleration noise spectrum which can be converted to strain via $\tilde{h}_{xx}(f)=-\frac{1}{x_0 \omega^2} \tilde{\ddot{x}}(f)$. The corresponding strain spectral density is shown in Figure \ref{StrainLPF}. This measurement can constrain the stochastic gravitational wave background to $\Omega_{GW}<6.5 \times 10^{8}$ at 10 mHz as seen in Figure \ref{GW}.

\end{document}